\documentclass{article}
\usepackage[dvips]{graphicx}
\usepackage[small]{caption}

\title{Phonon Signature of Charge Inhomogeneity in High Temperature Superconductors
YBa$_{2}$Cu$_{3}$O$_{6+x}$}

\author{{\normalsize Y. Petrov$^{1}$, T. Egami$^{2}$, R. J. McQueeney$^{3}$, M. Yethiraj$^{4}$,
H. A. Mook$^{4}$, and F. Dogan$^{5}$}\\ 
{\small\itshape $^{1}$Department of Physics and Astronomy, University of Pennsylvania, Philadelphia,}\\
{\small\itshape PA 19104, $^{2}$Department of Materials Science and Engineering, University of}\\ 
{\small\itshape Pennsylvania, Philadelphia, PA 19104, $^{3}$Los Alamos National Laboratory, Los Alamos,}\\
{\small\itshape NM 87545, $^{4}$Oak Ridge National Laboratory, Oak Ridge, TN 37831, $^{5}$Department of }\\
{\small\itshape Materials Sci-ence and Engineering, University of Washington, Seattle, WA 98195}}

\date{\today}

\begin{document}
\maketitle

\begin{abstract}
Temperature and composition dependences of high-energy longitudinal optical
(LO) phonons in YBa$_{2}$Cu$_{3}$O$_{6+x}$, studied by inelastic neutron
scattering measurements, provide clear evidence for temperature dependent
spatial inhomogeneity of doped charges. The measurements indicate that
charge doping increases the volume fraction of the charged regions, while
the local charge density remains unchanged. For the optimally doped sample
the charge distribution changes sharply near the superconducting transition
temperature, with stronger charge inhomogeneity below T$_{C}$. The
remarkable magnitude of the phonon response to charge suggests strong
involvement of phonons in charge dynamics.

\end{abstract}

For some time, high-temperature superconductivity in cuprates has been
believed to occur in a homogeneous system through a magnetic mechanism.
However, these assumptions are seriously challenged by recent experimental
observations that suggest spatial charge inhomogeneity and lattice effects.
In particular, the observation of the spin/charge stripe structure for
non-superconducting La$_{1.475}$Nd$_{0.4}$Sr$_{0.125}$CuO$_{4}$ has
eloquently explained the observed incommensurate magnetic and lattice
diffraction \cite{1}. In superconducting samples, the incommensurate
periodicity is observed only in the dynamic spin structure. The dynamic
periodicity changes almost linearly with composition and is consistent with
the static superlattice periodicity for the non-superconducting composition 
\cite{2,3}. Thus, it was postulated that dynamic stripe correlations
exist in superconducting cuprates \cite{1}, and various theories have been
advanced assuming such correlations \cite{4,5,6}. However,
direct evidence of a lattice signature of the charge inhomogeneity is
incomplete. While our earlier work on the LO phonons in La$_{1.85}$Sr$%
_{0.15} $CuO$_{4}$ (LSCO) implicated charge inhomogeneity \cite{7}, only one
composition was studied. Phonon anomalies were recently observed for YBa$%
_{2} $Cu$_{3}$O$_{6+x}$ (YBCO, x = 0.2, 0.35, and 0.6) \cite{8}, but
anomalies are rather weak, and temperature dependence was only cursorily
characterized. In this report, we describe the results of neutron inelastic
scattering measurements of high-energy LO phonons in superconducting YBCO (x
= 0.2, 0.35, 0.6, and 0.93) that not only provide strong evidence for
temperature dependent charge inhomogeneity, but also clarify important
features of the charge inhomogeneity. The results indicate that the local
charge density of the charged region remains constant with doping, even in
the optimally doped sample, and doping only changes the volume fraction of
the charged region. The size of the charged region is microscopic, and for
the optimally doped sample the charge separation is significantly enhanced
in the superconducting phase.

The studied YBCO samples were large single crystal disks with a height of
about 1 cm and diameter ranging from 5 cm for the optimally doped sample (x
= 0.93) to 4 cm for the strongly underdoped sample (x = 0.2). Measurements
were carried out with the HB-3 triple axis spectrometer at the High Flux
Isotope Reactor at Oak Ridge National Laboratory. To monochromatize incident
neutrons a beryllium (101) reflection was used, while a pyrolitic graphite
(002) reflection (Silicon (111) for x = 0.6 sample) was used for an
analyzer, set to give a final neutron energy of 14.7 meV. The angular
divergence of the beam was 48'-40'-80'-120' for large x = 0.93 crystal and
48'-60'-80'-240' for smaller x = 0.2, 0.35 and 0.6 crystals. As in Ref. 7 we
focused on the LO mode along the in-plane Cu-O bond direction which is polar
at the zone-center and half-breathing at the zone-edge. Inelastic neutron
scattering measurements were made at energy transfers from 50 - 80 meV and
momentum transfers, \textit{Q}, along the (100) direction from (3, 0, 0) to
(3.5, 0, 0) in the unit of the reciprocal lattice vector (\textit{a}* = 1.63
\AA $^{-1}$). Thus, the measurement only detects longitudinal optical (LO)
phonons. Since the samples are twinned, \textit{a}-axis and \textit{b}-axis
phonons are observed at the same time.

\begin{figure}[tb]
   \centering 
   \includegraphics[height=8cm]{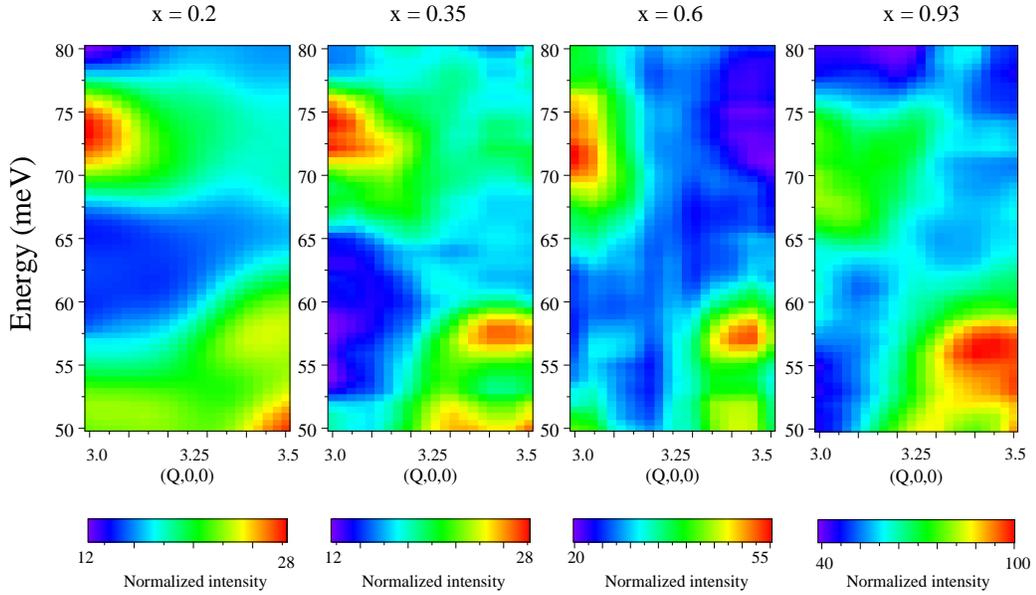}
   \caption{ Composition dependence of the inelastic neutron scattering
intensity from YBCO single crystals with x = 0.2, 0.35, 0.6 and 0.93, at T = 10 K.}
   \label{f1}
\end{figure}

The inelastic neutron scattering intensity pattern at T = 10 K for various
doping levels is shown in Fig.~\ref{f1}. In this system the undoped (x = 0) sample
has a dispersionless LO branch at 75 meV, while doping softens the zone-edge
mode down to 55 meV \cite{9}. Thus one might expect continuous softening at
the zone-edge as doping level is increased. Instead, Fig. 1 shows that the
LO branch is always split into two, the high-energy branch around 75 meV and
the low-energy branch around 55 meV. Instead of continuous softening the
spectral weight is transferred from the high-energy branch to the low-energy
one as doping is increased. This result is most naturally understood in the
two-phase picture if we associate the high-energy and low-energy branches
with the micro-phases possessing low and high charge densities,
respectively. When the charges are segregated into microscopic domains,
increasing the doping level does not change the local charge density in the
domains, but simply increases their total volume fraction, causing the
spectral weight transfer from the high-energy branch to the low-energy one.
The characteristic \textit{Q}-dependence of the two branches indicates that
the size of the charged domain is microscopic, since otherwise two branches
with a similar \textit{Q}-dependence in intensity will be observed.

For the optimally doped sample the inelastic magnetic neutron scattering is
dominated by the resonance at 41 meV, so that the evidence of incommensurate
magnetic modulation is weak \cite{10}. The present data, however, indicate
very clear and strong charge inhomogeneity for the x = 0.93 sample,
suggesting that the magnetic modulation may not be the dominant force to
produce charge inhomogeneity. This is in accord with the observation that
the charge signature sets in at a higher temperature than the spin signature
does for the sample with the 1/8 charge density, implying that charge is
driving the stripe ordering \cite{11}. The dispersion shown in Fig. 1 agrees
with the earlier result for the optimally doped YBCO \cite{9}, and has a
striking resemblance to the one obtained earlier for LSCO \cite{7}.

\begin{figure}[tb]
   \centering 
   \includegraphics[height=8cm]{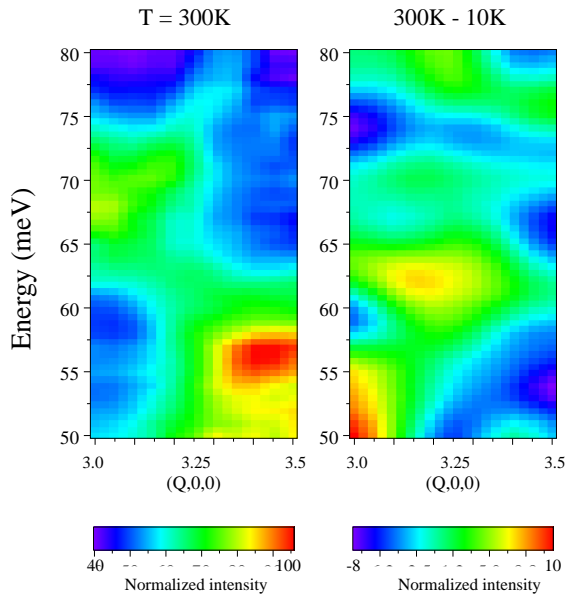}
   \caption{(a) Inelastic neutron scattering intensity map for the optimally
doped (x = 0.93) YBCO single crystal at T = 300 K, and (b) the difference
between the data at 300 K and 10 K.}
   \label{f2}
\end{figure}
Fig.~\ref{f2} shows the intensity pattern for the x = 0.93 sample obtained at room
temperature, \textit{I}$_{300K}$, and the difference, \textit{I}$_{300K}$ - 
\textit{I}$_{10K}$. At room temperature the two branches are more connected
at the middle, in agreement with the result on LSCO \cite{7}. In order to
study the temperature dependence in more detail the scattering intensity as
a function of energy transfer was measured at (3.25, 0, 0) at various
temperatures. The average intensities over the energy range from 56 to 68
\begin{figure}[tb]
   \centering 
   \includegraphics[angle=-90, totalheight=10cm]{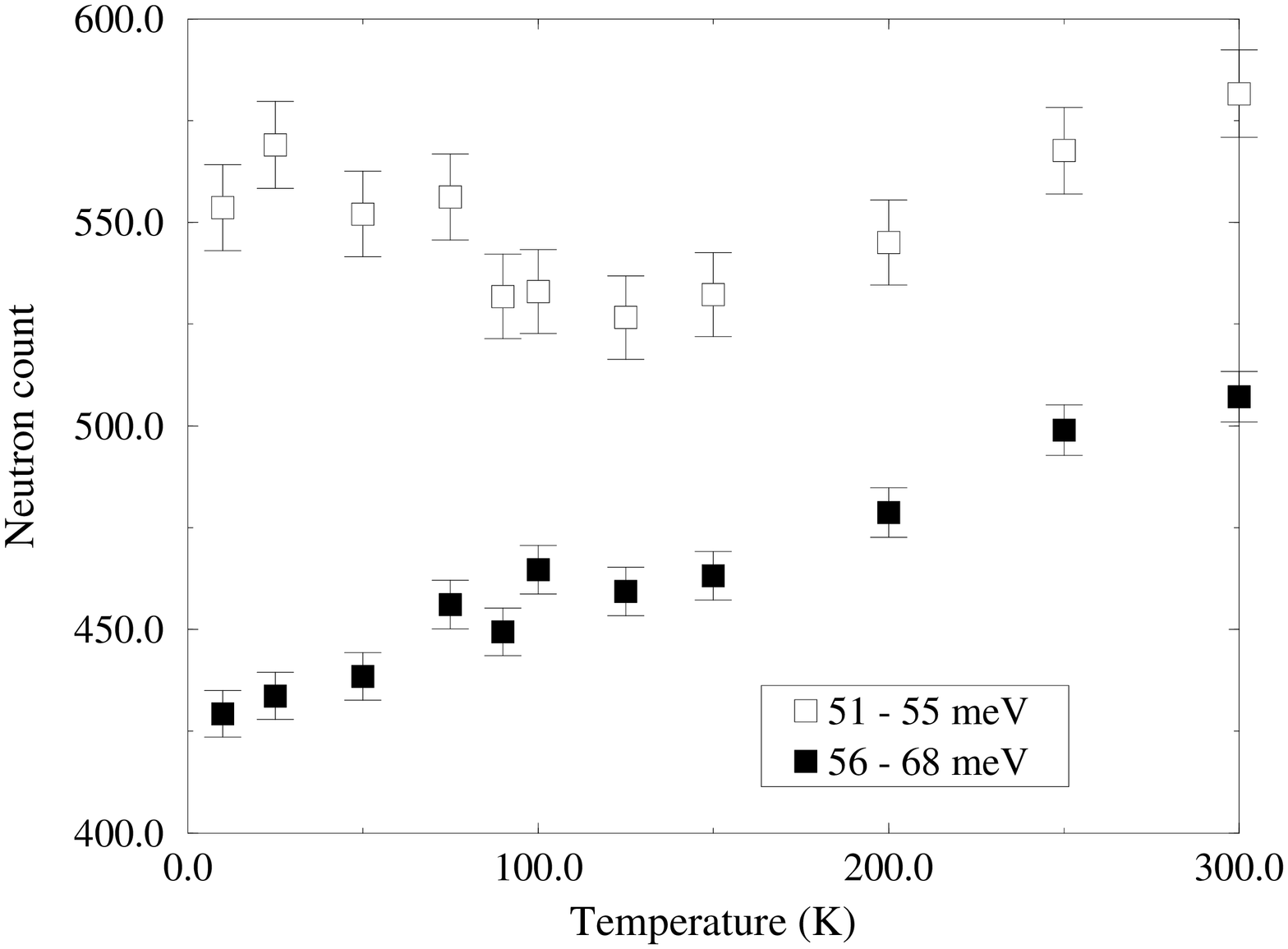}
   \caption{(a) Temperature dependence of the average neutron scattering
intensity at (3.25, 0, 0) over the regions of energy transfer from 56 to 68
meV, I(1), from 51 to 55 meV, I(2), and (b) I(2) – I(1).  The horizontal
line in (b) is a guide to the eye.  The superconducting transition
temperature, TC, is 95 K.}
   \label{f3}
\end{figure}
meV (Range 1), \textit{I}(1), and from 51 to 55 meV (Range 2), \textit{I}%
(2), are plotted in Fig.~\ref{f3}(a), and the difference, \textit{I}(2) - \textit{I}%
(1), in Fig.~\ref{f3}(b). It is clear that a sharp change is observed near the
superconducting transition temperature of T$_{C}$ = 95 K. Above T$_{C}$, 
\textit{I}(2) - \textit{I}(1) is practically constant. This indicates that
the spectral weight is shifted from Range 2 to Range 1 as temperature is
raised up to T$_{C}$, with no further change above T$_{C}$. Since for the
optimally doped YBCO T$_{C}$ coincides with the pseudo-gap temperature T$^*$,
it is not clear whether the change is associated with T$_{C}$ or with T$^*$.
Recent reports on the isotope effect on T$^*$ \cite{12,13} lead us to
think the latter is more likely. Our preliminary study on x = 0.35 indicates
gradual change and saturation around 200 K, which is close to T$^*$. An
objection was raised recently regarding the temperature dependence in LSCO 
\cite{14}, this issue involves the problem of focusing of the spectrometer,
and will be discussed elsewhere. For YBCO the presence of strong temperature
dependence is unquestionable as presented here.

\begin{figure}[tb]
   \centering 
   \includegraphics[angle=-90, totalheight=10cm]{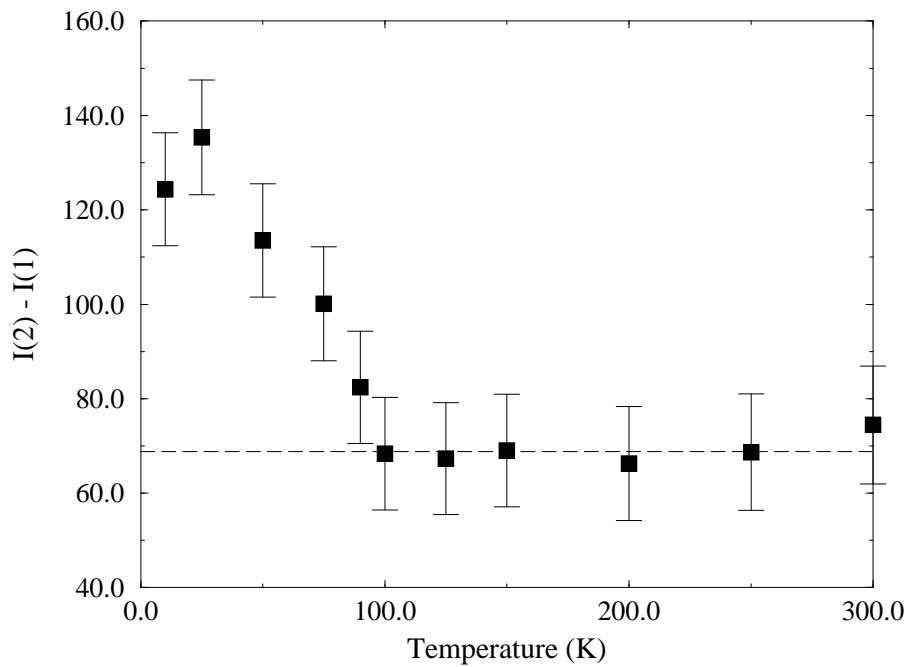}
   \caption{High-energy Cu-O LO mode, at the zone-center (above) and zone-edge
(below).}
   \label{f4}
\end{figure}

The results described here represent the most convincing evidence obtained
so far on the spatial inhomogeneity of the charged state in the
superconducting cuprates, including the optimally doped sample. While the
pattern of inhomogeneity is not directly discernible from the present
result, a simulation indicates that a quasi-static stripe pattern is not
compatible with the split dispersion, and the cell-doubling model exhibits a
better fit \cite{7}. But the shape of the domains is probably stripe-like,
since the size of the domain speculated from the shape of the
dispersion-less portion of the phonon dispersion is 8 $\times $ 20 \AA\ \cite
{7}. Thus it is possible to characterize the charged domains as fragments of
stripes. The fact that greater charge inhomogeneity was observed below T$%
_{C} $/T$^*$ implies that charge inhomogeneity does not compete against
superconductivity, but may facilitate its occurrence. As shown in Fig.~\ref{f4} the
half-breathing mode at the zone-edge results in charge transfer between Cu
and O, while the polar mode at the zone-center does not. Thus the zone-edge
mode is expected to couple strongly to the doped holes in a charge transfer
system such as the cuprates \cite{15,16,17}. Indeed the two
branches are remarkably different in energy (15 - 20 meV), suggesting strong
involvement of the low-energy phonon branch in charge dynamics.

Such microscopic inhomogeneity was observed for manganites that show
colossal magnetoresistance (CMR), even in the metallic phase \cite{18,19}. Also strong phonon softening similar to the one shown here was
observed for manganites \cite{20}, suggesting intimate involvement of
phonons in producing charge segregation. At low doping levels charges in
manganites are localized as spin/lattice polarons or polaron aggregates, but
as the doping level is increased the polaron aggregates percolate to produce
an insulator-to-metal transition \cite{18,19}. In the metallic phase,
charges in the percolating charged domains are not localized, but still
spatially restricted. It is possible that a similar picture applies to
cuprates, probably with a smaller length scale. While the details of the
charged domains and the pairing mechanism are still unclear, the present
results strongly challenge conventional wisdom concerning the origin of
high-temperature superconductivity.

The authors are grateful to A. R. Bishop, V. J. Emery, L. P. Gor'kov, A.
Bussmann-Holder, and V. Kresin for useful discussions. The research at
University of Pennsylvania was supported by the National Science Foundation
through DMR96-28136. Measurements were made at Oak Ridge National
Laboratory, which is managed by Lockheed Martin Energy Research under
contract no. DE-AC05-96OR22464 for the Department of Energy.

\end{document}